# Helium Suppression in Impulsive Solar Energetic-Particle Events

**Donald V. Reames**

Institute for Physical Science and Technology, University of Maryland, College Park, MD 20742-2431 USA, email: dvreames@umd.edu

**Abstract** We have studied the element abundances and energy spectra of the small "He-poor" impulsive solar energetic-particle (SEP) events, comparing them with other impulsive SEP events with more-normal abundances of He. He-poor events can have abundances as low as He/O ≈ 2, while both impulsive and gradual SEP events usually have source abundances of 30 ≤ He/O ≤ 100 with mean values of ≈ 50 – 60. He/C ratios are not only low, but often decrease with energy in He-poor events. Abundance enhancement patterns of other elements with atomic numbers 6 ≤ $Z$ ≤ 56, and likely values of their mass-to-charge ratios *A/Q*, are generally unaltered in He-poor events, as are the probable source-plasma temperatures of 2.5 – 3.2 MK for all impulsive SEP events. One He-poor event is also an example of a rarer C-poor event with C/O = 0.08 ± 0.04, suppressed by a factor over 5 from the mean. We discuss suggestions of a possible *A/Q* threshold during acceleration and of the sluggish ionization of He entering the corona, because of its uniquely high first ionization potential (FIP), but the suppression of He and the decline of He/C with energy is difficult to explain if both He and C are fully ionized with *A/Q* = 2 as expected at 2.5 – 3.2 MK. Although less dramatic, a possible excess enhancement of Ne in some impulsive SEP events is also considered. Possible causes of the large ≈30% spectral and abundance variations in impulsive events are also discussed. However, the physics of the He-poor events remains a mystery.

Keywords: Solar energetic particles · Solar system abundances · Solar flares

## 1. Introduction

Studies of solar energetic particles (SEPs) have identified two distinct physical mechanisms of particle acceleration (Reames, 1988, 1999, 2013, 2015, 2017a; Gosling, 1993). The smaller "impulsive" SEP events have extreme 1000-fold enhancements of $^3$He/$^4$He (Serlemitsos and Balasubrahmanyan, 1975; Reames, von Rosenvinge, and Lin, 1985; Mason, 2007; Reames, 2017a) and of and heavy elements up to Os – Pb (Reames, 2000, 2017a; Mason *et al.*, 2004; Reames and Ng, 2004; Reames, Cliver, and Kahler, 2014a) and are associated with solar jets (Kahler, Reames, and Sheeley, 2001; Bučík *et al.*, 2018), with magnetic reconnection (Drake *et al.*, 2009), and with wave-particle resonance (Temerin and Roth, 1992; Roth and Temerin, 1997).

In contrast, SEPs in the larger gradual SEP events (Desai and Giacalone, 2016; Reames, 2017a) are accelerated at shock waves (Lee 1983, 2005; Zank, Rice, and Wu, 2000; Cliver, Kahler, and Reames, 2004; Ng and Reames, 2008; Rouillard *et al.*, 2012; Lee, Mewaldt, and Giacalone, 2012) driven out from the Sun by wide fast coronal mass ejections (CMEs; Kahler *et al.*, 1984). These shock waves sample the ambient coronal plasma as well as suprathermal ions left over (Mason, Mazur, and Dwyer, 2002) from multiple impulsive SEP events (Reames, 2016b) especially in active regions where small impulsive events occur regularly (Bučík *et al.*, 2014, 2015, 2018; Chen *et al.*, 2015).

The relative abundances of chemical elements in the SEPs differ from solar abundances at the photosphere in two ways that have been recognized (*e.g.* Meyer, 1985). First, the transport of elements up into the corona depends upon the first ionization potential (FIP) of the element. Low-FIP (< 10 eV) elements that are ionized in the photosphere can be preferentially boosted up into the corona by Alfvén waves (Laming, 2015), while high-FIP neutral atoms cannot. All SEP events seem to have the same underlying FIP dependence (Reames, 2018b) which differs from that in the solar wind (Mewaldt *et al.*, 2002; Desai *et al.*, 2003; Schmelz *et al.*, 2012; Reames, 2018a). Second, element abundances in individual SEP events, or individual time periods within events, can show power-law dependence on the mass-to-charge ratio *A/Q* of the ions, both for impulsive (Reames, Meyer, and von Rosenvinge, 1994; Mason *et al.*, 2004; Reames and Ng, 2004;

Reames, Cliver, and Kahler, 2014a, 2014b) and for gradual (Meyer, 1985; Breneman and Stone, 1985; Reames, 2016a, 2016b, 2018b) SEP events. A power-law dependence of ion scattering upon magnetic rigidity during acceleration or transport (Parker, 1963; Ng, Reames, and Tylka, 1999, 2001, 2003, 2012; Reames, 2016a) becomes a power-law dependence upon $A/Q$ when ion abundances are compared at the same particle velocity.

In the scattering during transport of ions from gradual SEP events, Fe, for example, with higher $A/Q$, scatters less than O, so that Fe/O is elevated early in an event and depleted later. This time variation results from a spatial variation in Fe/O that is then further modified by solar rotation. Averaging over many times in many events tends to wash out these spatial variations and helps to recover the original abundances: the reference SEP coronal abundances (Meyer, 1985; Reames, 1995, 2014, 2017a). These may be compared with the theory (*e.g.* Laming, 2015) of the FIP-effect and with other measures of coronal abundances such as the solar wind where the pattern of the FIP-effect differs (*e.g.* Bochsler, 2009; Schmelz *et al.*, 2012; Reames, 2018a, 2018b, 2018c). These reference abundances from gradual SEP events also form a basis for the study of impulsive SEP events where the strong $A/Q$ dependence is probably produced during acceleration in islands of magnetic reconnection in the source (*e.g.* Drake *et al.*, 2009). The reference SEP abundances are listed in Appendix A. We use enhancement ≡ observed / reference.

An early study of element abundances in impulsive SEP events (Reames, Meyer, and von Rosenvinge, 1994) noted groupings of abundances where He, C, N, and O were unenhanced relative to the reference abundances, Ne, Mg, and Si were enhanced by a factor of ≈ 2.5, and Fe was enhanced by a factor of ≈ 7. The groupings suggested that C, N, and O were nearly fully ionized, like He, with $A/Q$ ≈ 2, while Ne, Mg, and Si were in an energetically stable state with two orbital electrons. This occurs at an electron temperature of $T$ = 3 – 5 MK. At lower temperatures O would begin to pick up electrons and at higher temperatures Ne would soon become fully ionized like He. More recently Reames, Cliver, and Kahler (2014a) prepared a list on 111 impulsive SEP events occurring over nearly 20 years, and Reames, Cliver, and Kahler (2014b) developed a best-fit scheme to determine the source temperature and power-law fit of abundance enhancements *versus* $A/Q$ for each event. Each temperature implies $Q$ and $A/Q$ values for each element and a fit of enhancement *versus* $A/Q$ gives a $\chi^2$ value for that $T$. For each impul-sive SEP event we choose the fit and the value of $T$ with the minimum $\chi^2$. Nearly all of the measurements were found to fall in the 2.5 – 3.2 MK temperature range (Reames, Cliver, and Kahler, 2014b, 2015), but the scatter of the points around the fit line was found to be 20 – 30%, larger than expected from the statistical errors alone.

A similar study of the source temperature and the power-law fit *versus A/Q* for gradual SEP events was carried out by Reames (2016a, 2016b) where the range of $T$ was found to be larger, from 0.8 to 3.2 MK. However, when T > 2.5 MK, the variations in He, C, and O were much smaller in gradual SEP events than in impulsive ones (Reames, 2016b, 2018b), a result of greater averaging. More generally, the source values of He/O are found to vary rather widely from 30 to 120 (Reames, 2017b, 2018b), compared with previous reference values of 47 (Reames, 2014) or 57 (Reames, 1995). It is also true that those gradual SEP events with $T \geq 2$ MK have higher source He/O $\geq 60$ (Reames, 2017b, 2018b); those are the events where the shock wave most likely reaccelerated residual suprathermal ions from prior impulsive SEP events (Tylka *et al.*, 2005; Tylka and Lee, 2006).

What about impulsive SEP events? Are the variations in He consistent with variations of other elements about the best-fit power law? Why are all of these variations so large? Are they energy dependent? Are the variations dependent on features of the acceleration model, on the FIP process, or on both? How do He-poor events differ?

In this paper we revisit the 111 impulsive SEP events on the list given by Reames, Cliver, and Kahler (2014a). Event properties including onset times and durations are included in that list. We briefly demonstrate the fitting procedure and then show typical power-law fits and energy dependences for events with well-behaved He abundances. These are then contrasted with the power-law fits and energy-spectral behavior of several "He-poor" events, a "C-poor" event, and two "Ne-rich" events. We use observations made by the Low-Energy Matrix Telescope (LEMT) on the *Wind* spacecraft, near Earth (von Rosenvinge *et al.*, 1995; see also Chapter 7 of Reames, 2017a). LEMT primarily measures elements He through Pb in the region of 2 – 20 MeV amu$^{-1}$. The element resolution of LEMT up through Fe is shown in detail by Reames (2014). LEMT resolves element groups above Fe as shown by Reames (2000, 2017a).

Throughout this text, whenever we refer to the element He or its abundance, we mean $^4$He, unless $^3$He is explicitly stated.

## 2. Fitting Powers of *A/Q*

Increasingly, the key to understanding the relative enhancements of the elements has been the correlated pattern of their *A/Q* values as their *Q* values vary with source temperature. The dependence of *A/Q* on *T* is shown in Figure 1. The figure shows the grouping pattern of major elements found by Reames, Meyer, and von Rosenvinge (1994) that occurs near 3.2 MK and shows a region where $(A/Q)_{Ne} > (A/Q)_{Mg} > (A/Q)_{Si}$ near 2.5 MK; a similar pattern of enhancements is often seen.

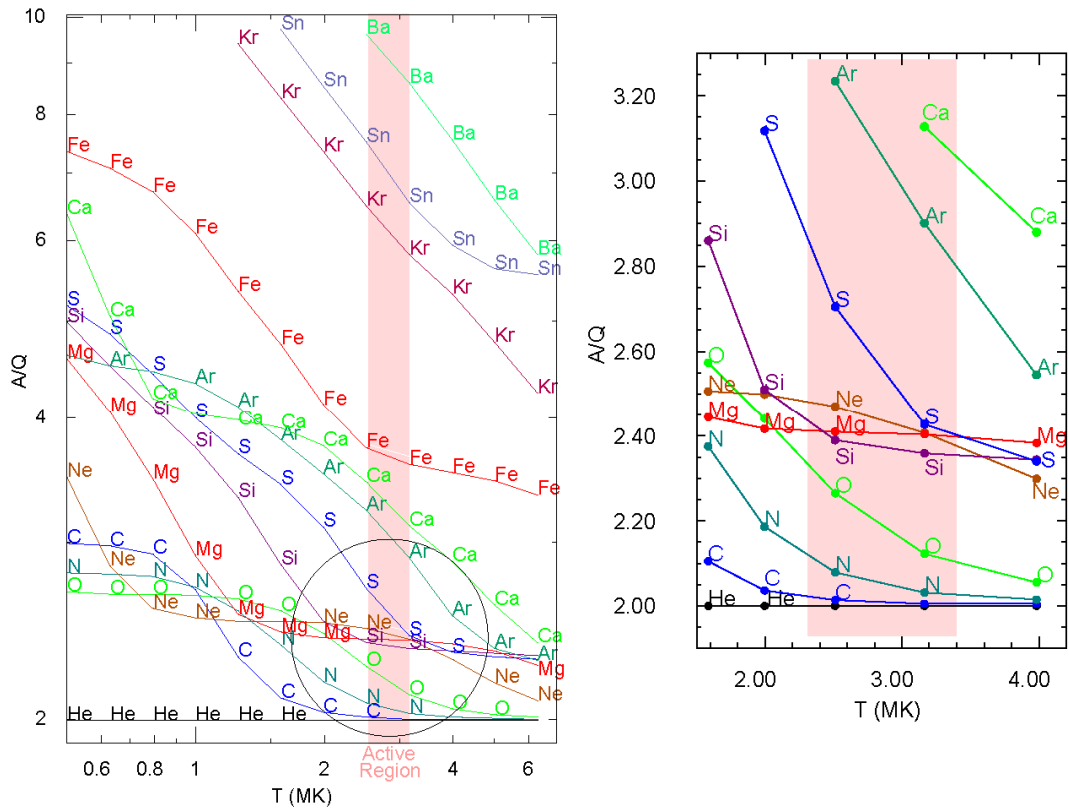

**Figure. 1**. The dependence of *A/Q* on *T* for selected elements as indicated. Values from He through Fe are from Mazzotta *et al.* (1998); those for higher *Z* are from Post *et al.* (1977). The region circled in the *left panel* is magnified in the *right panel* especially to show the ordering of Ne, Mg, Si, and S in the 2.5 – 3.2 MK region. The ordering of abundance enhancements follows the ordering of *A/Q* at the source *T*.

A typical analysis of the best-fit *A/Q* power law and temperature is shown for a series of three impulsive SEP events in Figures 2 and 3. In the lower panel of Figure 2, the event onsets are labeled 61, 62, and 63, corresponding to their numbers in the list of Reames, Cliver, and Kahler (2014a). For each event, the element enhancements relative to reference abundances are plotted and fit *versus A/Q* values for each temperature we study, and $\chi^2$ for the fit is plotted *versus T* as shown in the upper-right panel of Figure 2. The minimum value of $\chi^2$ determines the best-fit temperature (shown in the upper-left panel of Figure 2) and the best-fit power law *versus A/Q* (shown in Figure 3).

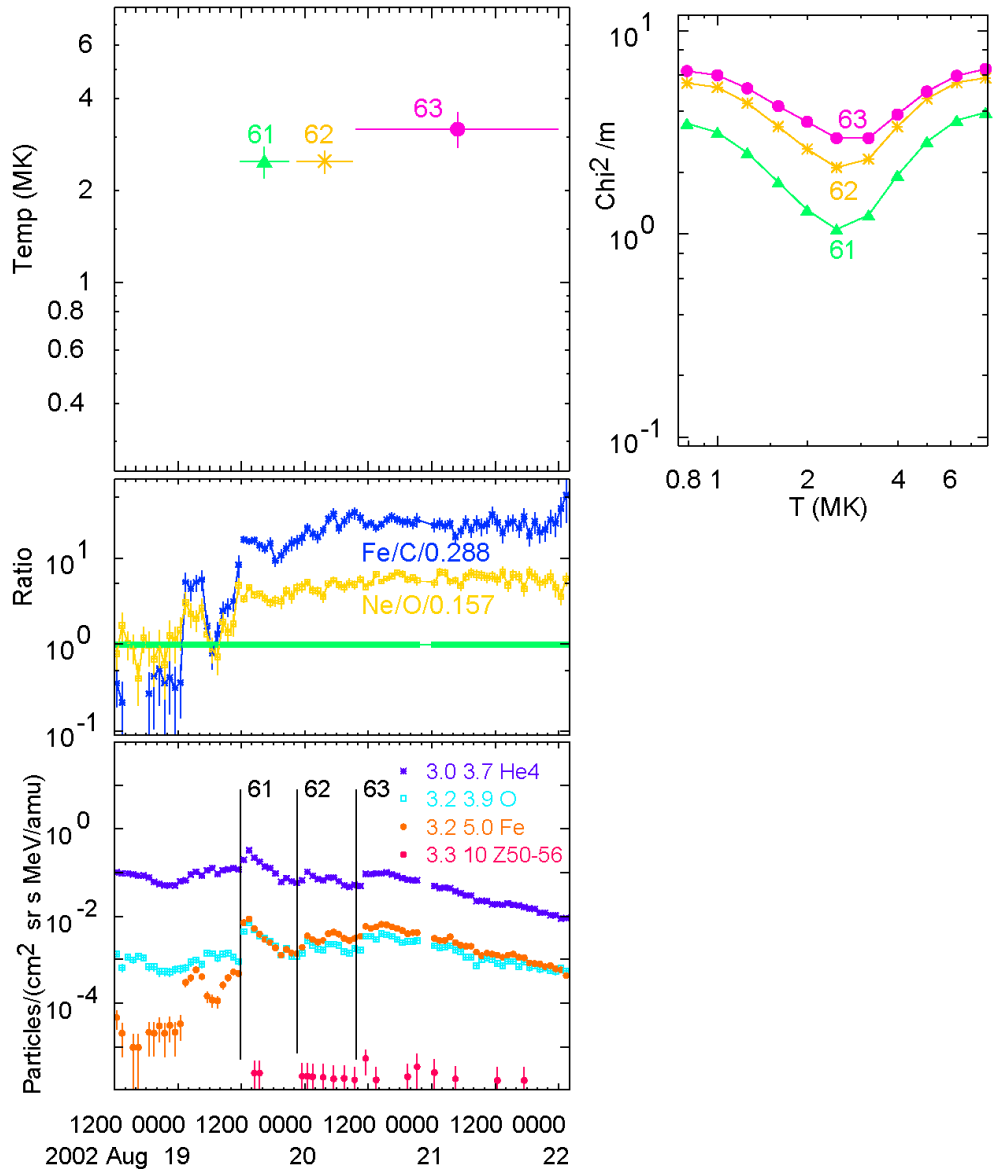

**Figure 2**. Time profiles for $^4$He, O, Fe, and $50 \leq Z \leq 56$ for events 61, 62, and 63 are shown in the *lower panel*, normalized ratios of Fe/C and Ne/O in the *middle panel* and the best-fit temperatures in the *upper-left panel*. $\chi^2$ *versus T* is shown for each event in the *upper-right panel*.

**Figure 3**. The best-fit power-law dependence of abundance enhancement *versus* *A/Q* is shown for each of the three impulsive SEP events shown in Figure 2. The atomic number of the element, *Z*, is listed at each data point. The *A/Q* values are based upon *Q* values at the best-fit temperature at the $\chi^2$ minimum shown in the *upper-right panel* of Figure 2.

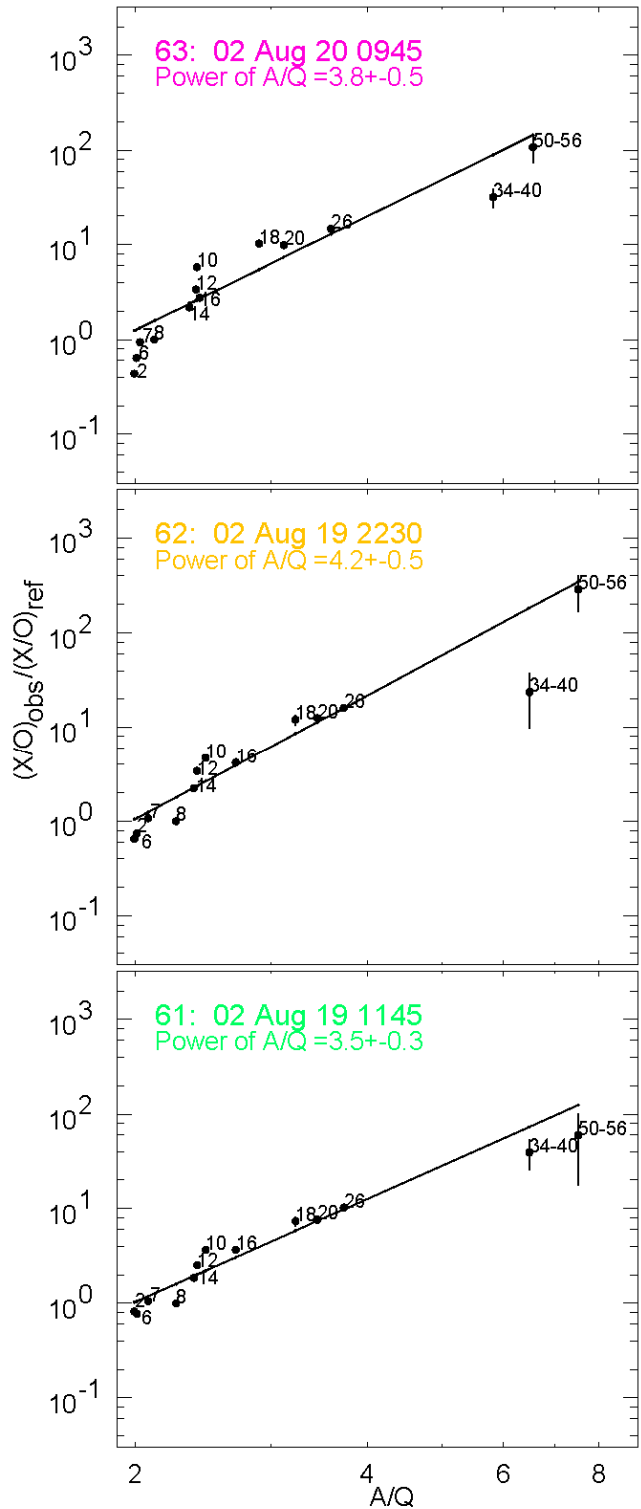

In Figure 3, the enhancements of He, C, N, and O tend to group together, and the enhancements of Ne, Mg, and Si are above them with Ne > Mg > Si in all three cases. Enhancements show S < Ne and S < Mg, in event 63; this lower value of enhancement for S tends to favor $T = 3.2$ MK for this event but $T = 2.5$ MK for the earlier events.

In the figure, the heavier elements such as $50 \leq Z \leq 56$ fit reasonably well despite poorer statistics. This is better illustrated for averages of impulsive SEP events where the improved statistics show a consistent power law up to $78 \leq Z \leq 82$ at $T \approx 3$ MK while also supporting the complex abundance pattern near Ne (Reames, Cliver, and Kahler, 2014a). We do not include the region $78 \leq Z \leq 82$ in fitting individual events because of its extremely low abundances.

## 3. Typical Impulsive SEP Events

In order to determine deviant behavior we must exhibit the properties of typical impulsive SEP events, especially with respect to their He abundances. Figure 4 shows enhancement *versus A/Q* for six representative impulsive SEP events that are not He-poor. In each case, the $Q$ values for each event were determined at the best-fit temperature as found from curves of $\chi^2$ *versus T* like the examples shown in the upper-right panel of Figure 2.

To begin a study of He abundances, plots of He/C *versus* energy for the same six events are shown in Figure 5. Observed values of He/C are used directly since there is no power-law correction to these abundances if He and C are both fully ionized with $A/Q = 2$ as shown in the fits in Figure 4. C is a better denominator than O for this reason even though ratios to O are usually quoted. Using the reference values of He/O = 57 ± 5 and C/O = 0.42 ± 0.01 (determined from averaging gradual SEP events and listed in Appendix A), we would expect He/C = 136 ± 13.

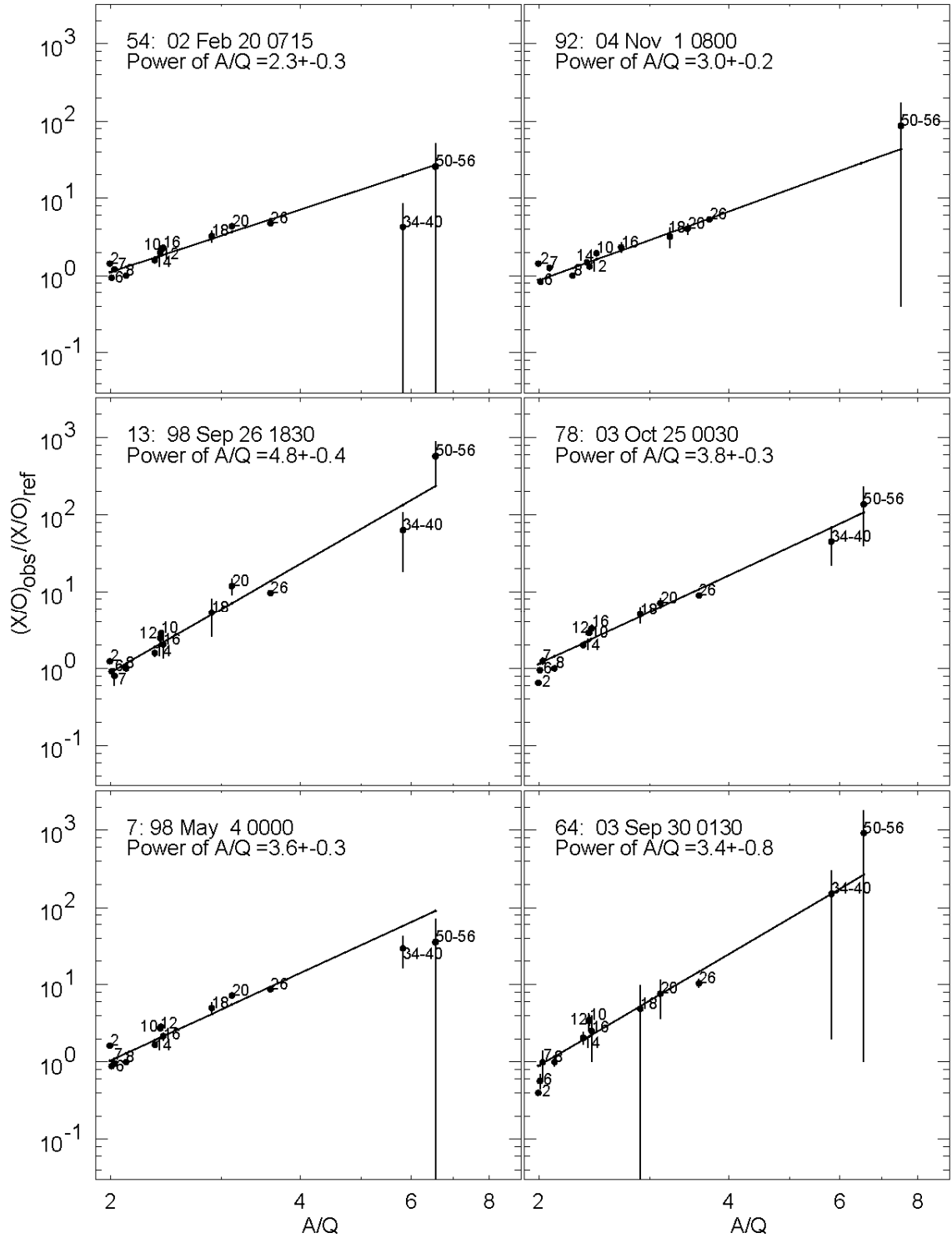

**Figure 4**. Abundance enhancements of elements, labeled by their atomic numbers $Z$, are plotted *versus* their best-fit $A/Q$ values for six typical impulsive SEP events, together with the best-fit least-squares fits showing their power-law behavior. Event numbers in each panel correspond to those in the list of Reames, Cliver, and Kahler (2014a)

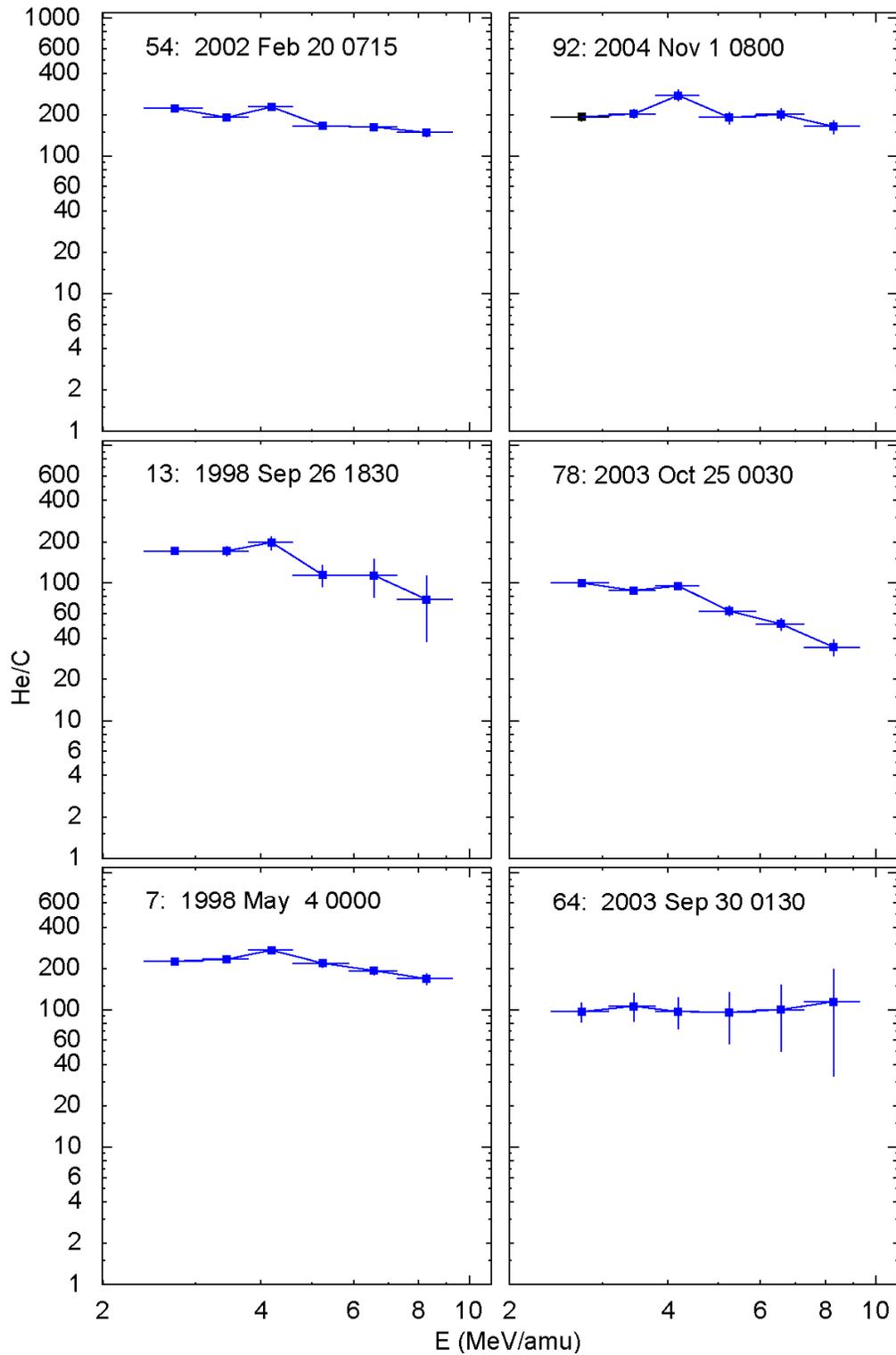

**Figure 5.** Observed values of He/C *versus* energy are shown for the same six impulsive SEP events studied in Figure 4. Event numbers (from Reames, Cliver, and Kahler, 2014a) and onset times are shown in each panel.

A histogram of the observed He/C ratios at 3 – 4 MeV amu$^{-1}$ for all of the impulsive SEP events from the list (Reames, Cliver and Kahler 2014a) is shown in Figure 6. The mean of the distribution is 126 ± 7. This corresponds to a source ratio of He/O = 53 ± 3 if we use the reference value of C/O. The mean and the statistical spread of individual measurements are shown in the figure. (These values use statistical weighting; the un-weighted mean of He/C is 115 ± 8.) As noted above, the average reference value of He/C determined from gradual SEP events (Reames, 1995, 2014, 2017a) is 136 ± 13. Statistically, the two samples are similar.

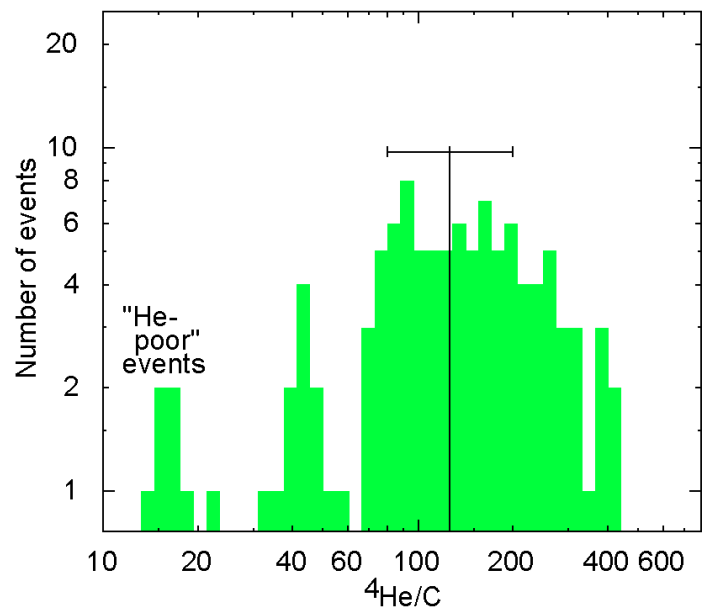

**Figure 6.** A histogram of the observed distribution of He/C at 3 – 4 MeV amu$^{-1}$ is shown for all of the impulsive SEP events. He-poor events are noted.

However, a unique feature of these impulsive SEP events is the "He-poor" events which have no analog in gradual SEP events. These He-poor events are small and thus have little effect on the overall accounting of He in the corona sampled by SEPs. In the next section we focus on the abundances, energy spectra, and other properties of these small but challenging events.

## 4. He-poor Events

Figure 7 shows the time dependence and energy spectra of He, O, and Fe, and the energy dependence of He/C, as well as the fit of element enhancements *versus A/Q* for event 79 of 30 December 2003. Clearly, the He enhancement, noted by the arrow, falls well below the fit of the other elements in the upper panel of Figure 7, although this is a small event and several elements, S, Ar, and Ca, cannot be measured. We will revisit the choice of the source temperature and the fitting parameters after more He-poor events have been shown.

The energy spectra and He/C ratio in Figure 7 show that the abundance ratios, He/C, He/O, and Fe/O, are not confined in energy, but are normalization factors that seem to apply to the whole event. He/C is much lower here than for the typical events in Figure 5. In addition, this event has $^3$He/$^4$He > 1, possibly reflecting the decrease in $^4$He.

Figure 8 shows parameters for the He-poor event of 27 June 2004 using the same format as Figure 7. This event has a high background of low-energy He prior to the event that might have contributed to the low-energy spectrum and He/C ratio. However, the measured He/C ratio is only ≈ 20 as it is, and would be even lower with any background subtracted.

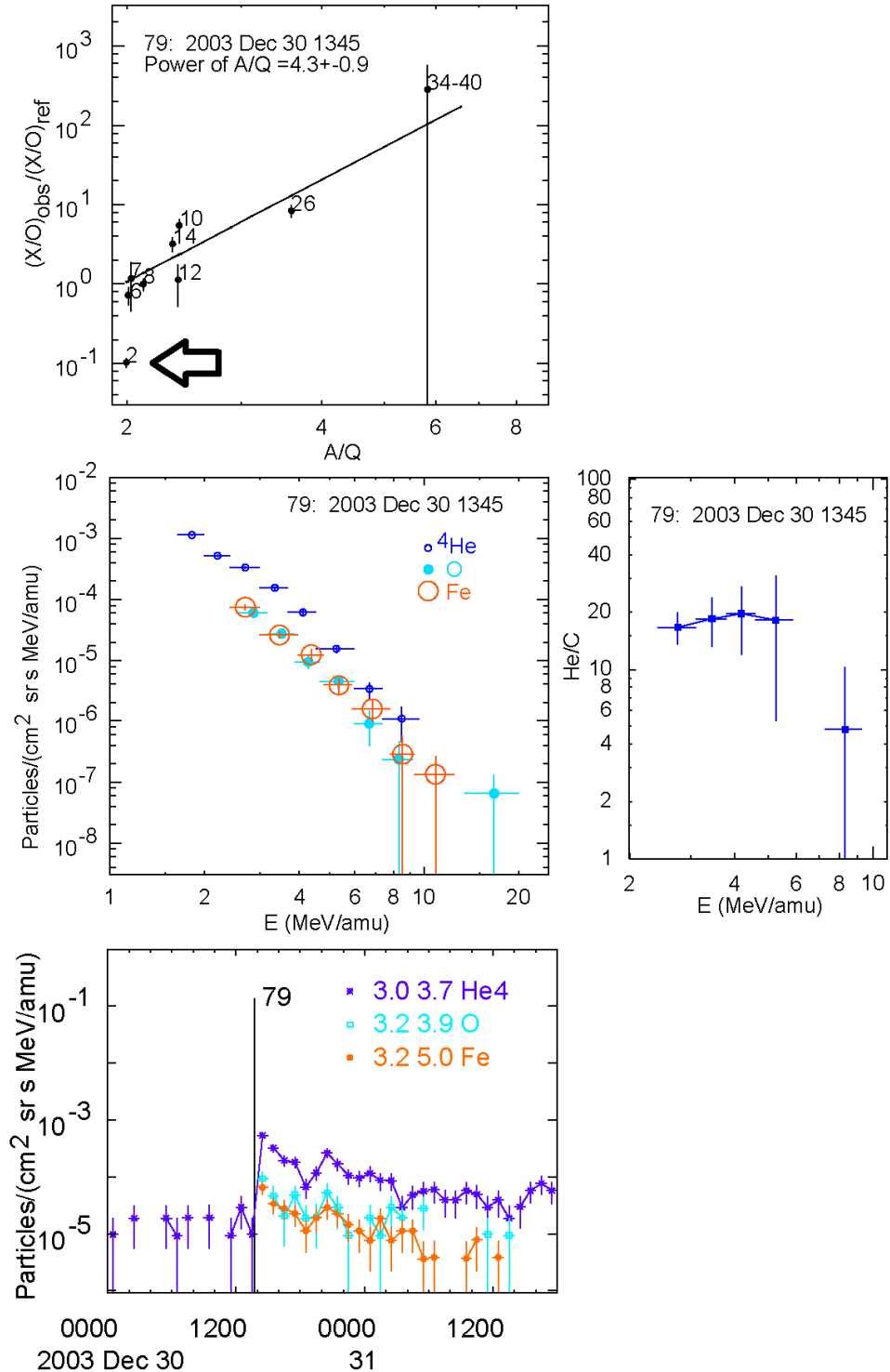

**Figure 7**. The time dependences of He, O, and Fe are shown in the *lower panel* for the He-poor impulsive SEP event 79 beginning at 1345 UT on 30 December 2003. A fit of element enhancements (labeled by *Z*) *versus A/Q* is shown in the *upper panel*, with depressed He noted by the arrow, while energy spectra of He, O, and Fe (*left*) and of He/C (*right*) are shown in the *center*.

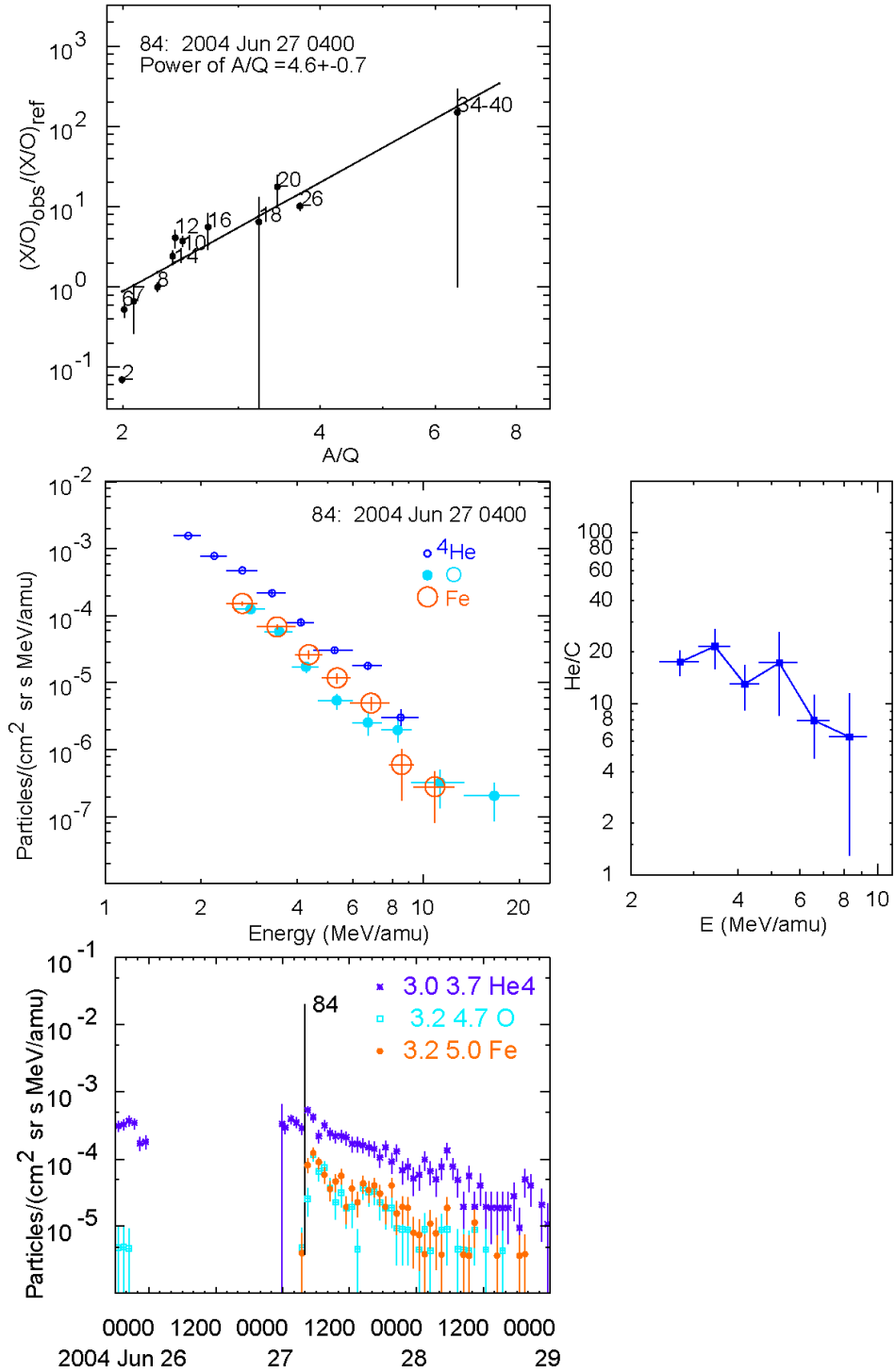

**Figure 8.** The time dependences of He, O, and Fe are shown in the *lower panel* for the He-poor impulsive SEP event 84 beginning at 0400 UT on 27 June 2004. A fit of element enhancements (labeled by *Z*) *versus* A/Q is shown in the *upper panel*, while energy spectra of He, O, and Fe (*left*) and of He/C (*right*) are shown in the *center*.

Figure 9 shows enhancements, spectra, and He/C for a pair of events in March 2000. Each event is preceded by some low-energy He background, although it does not seem adequate to explain the steeper He spectra and the He/C ratio that declines with energy. He in event 34 rises by a factor > 5 above pre-event background near 3 MeV amu$^{-1}$, suggesting a correction no larger than 20%. Note that event 35 has He/C ≈ 4 and He ≈ O ≈ Fe near 8 MeV amu$^{-1}$. This event is extremely He-poor. We have added Fe/O *versus E* to show that it differs from He/C.

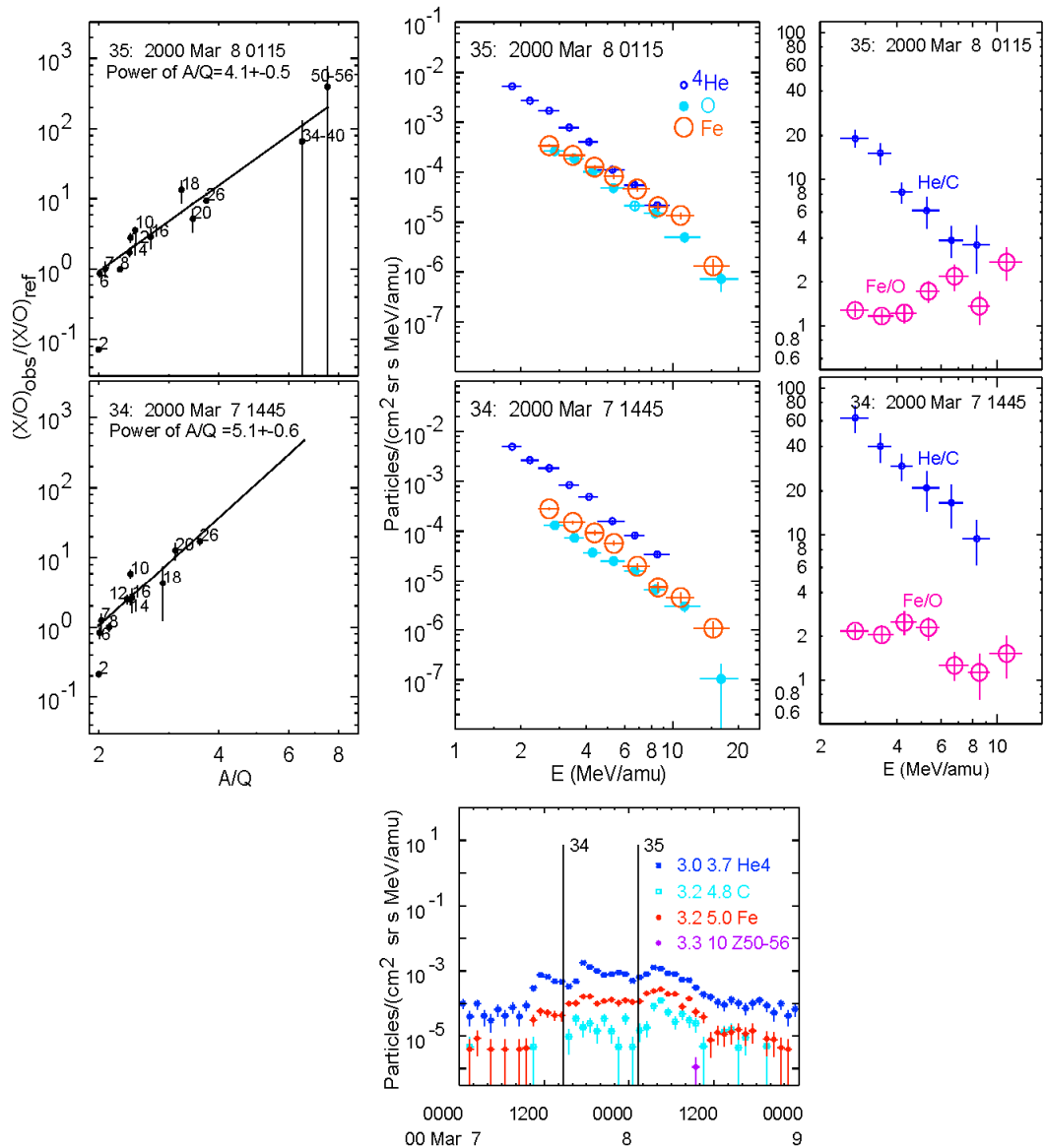

**Figure 9.** The *lower panel* shows the time history for a pair of events with low He/C ratios in March 2000. *Above* are shown the element abundance enhancement (labeled by *Z*) *versus A/Q* (*left*), energy spectra of He, O, and Fe (*center*), and ratios He/C and Fe/O (*right*) for each event listed in the panels.

Finally, we examine an impulsive SEP event that is both He-poor and C-poor in Figure 10. While the other elements follow the enhancement pattern of other impulsive SEP events, He and C fall well below the trend line. The low abundance of C is also clear in the lower panel of Figure 10. Here C/O = 0.08 ± 0.04.

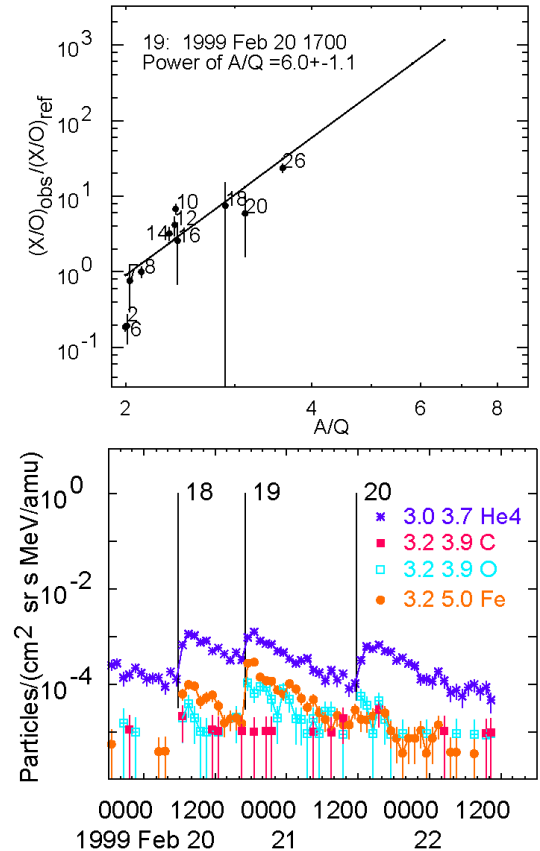

**Figure 10**. The *lower panel* shows intensities of He, C, O, and Fe for a series of three events with onset times flagged with event numbers 18, 19, and 20 in February 1999. The *upper panel* shows element abundance enhancements (labeled by *Z*) *versus A/Q* for the middle event, event 19. Here both He and C fall below the trend of the other elements, by about the same factor.

It is clear from Figure 1 that at temperatures of ≈ 1 MK, the *A/Q* value for He remains well below that of the heavier elements. At 1.5 MK, He and C lie together below the other elements. Yet we have not interpreted the He-poor events as a source-temperature variation. The reason is that the abundances of the elements with $Z > 6$ do not support this interpretation. At the lower temperatures, N and O rise to join and even exceed Ne and Mg, while Si and S leave the abundance group with Ne and rise to join Ar and Ca (see Figure 1). Such an abundance pattern at a source temperature of 1.5 MK is seen in some gradual SEP events (see Figure 7 in Reames, 2016a). However, the abundance pattern in most impulsive SEP events remains quite consistent: enhancements in Si and S usually fall below those in Ne and Mg. The enhancement pattern of the ele-ments N, O, Ne, Mg, Si, S, and Fe seems appropriate for 2.5 – 3.2 MK and He simply falls well below the trend line. Thus He is not suppressed because we have chosen the wrong source plasma temperature. However, even reducing $T$ to 1 MK would only reduce He/O by a factor of $\approx$ 3, not by an order of magnitude as observed.

He-poor impulsive SEP events rarely exceed intensities of $10^{-3}$ He ions (cm$^2$ sr s MeV amu$^{-1}$)$^{-1}$ at 3 MeV amu$^{-1}$, while many impulsive SEP events 10 – 100 times larger.

## 5. Ne-rich Events?

Impulsive SEP events can be reliably identified by elevated values of Fe/O and Ne/O in the region of a few MeV amu$^{-1}$ (see Figure 1 of Reames, Cliver, and Kahler, 2014a). This is largely because of the strong $A/Q$ dependence which enhances both abundances. One reason Ne/O stands out over Mg/O and Si/O is because of the higher value of $A/Q$ for Ne around 2.5 MK (see Figure 1). However, Ne/O often rises above the best-fit power law while Mg/O and Si/O are more evenly distributed. Is there an excess of Ne?

Figures 11 and 12 show two examples of events where the enhancement of Ne is a factor of $\approx$ 2 above the fitted power-law. The figures show spectra of O, Ne, and Si in these events and compare the Ne/O and Si/O ratios as a function of energy.

In event 44 on 11 July 2000 in Figure 11, the spectra are not pure power laws; they steepen somewhat with increasing energy. In addition, at 6 and 8 MeV amu$^{-1}$, O seems to rise above its trend while Si is depressed near 8 MeV amu$^{-1}$. This causes a dip in both Ne/O and Si/O near 8 MeV amu$^{-1}$.

In event 37 on 1 May 2000, shown in Figure 12, the Ne and O spectra are nearly identical, except near 8 MeV amu$^{-1}$ where the Ne intensity falls all the way down to that of Si. This produces a sharp drop in Ne/O while Si/O remains quite flat. Events with elevated Ne/O are not also He-poor. Other aspects of this well-known impulsive SEP event have been studied elsewhere (*e.g.* Reames, Ng, and Berdichevsky, 2001; Reames, 2018b).

While we cannot exclude the possibility of an added component of Ne with a steep spectrum (which we will discuss below), spectral variations of 20 – 30% in all species could easily explain the 20 – 30% fluctuations in abundance enhancements that seem to exist throughout.

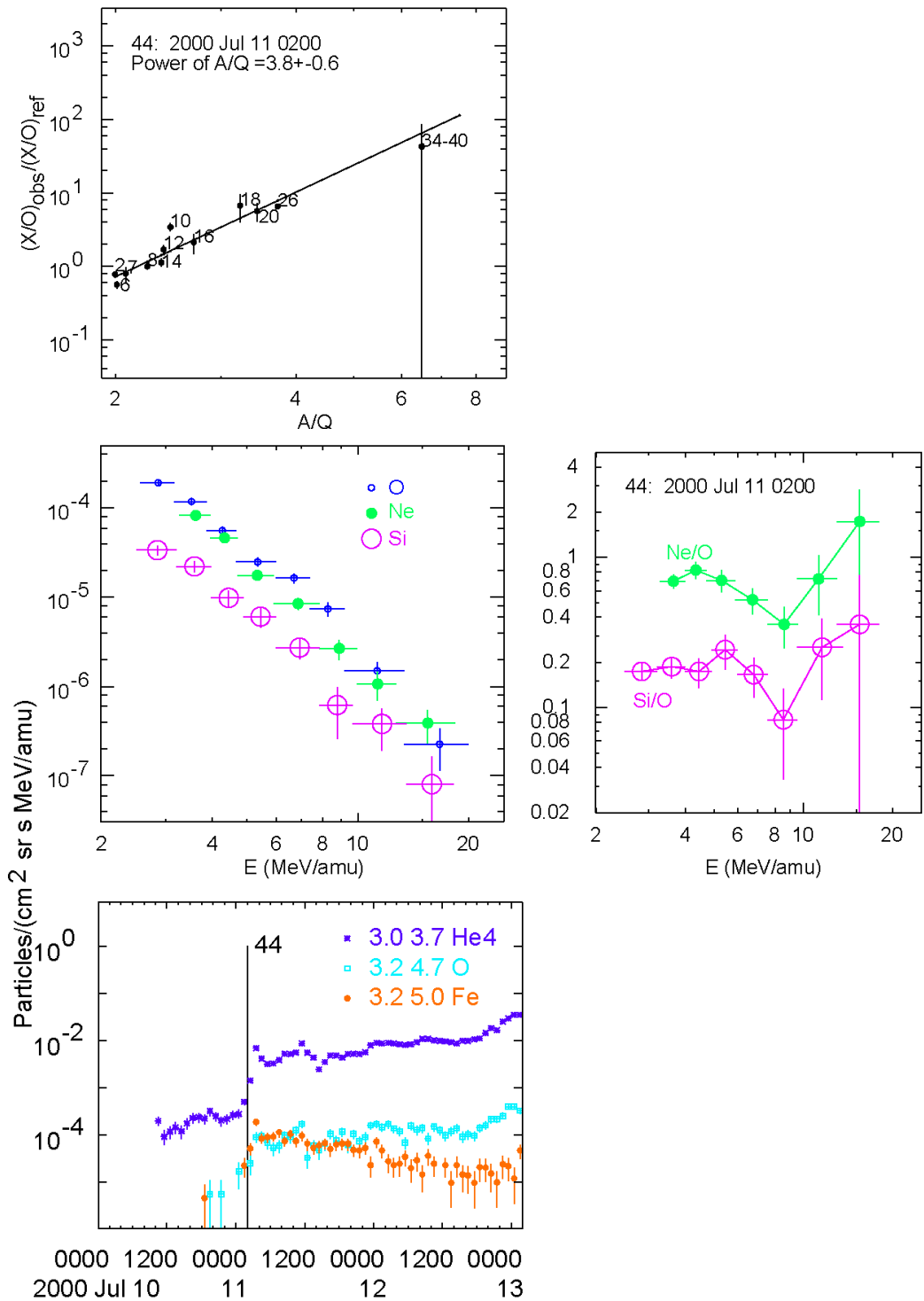

**Figure 11.** The *lower panel* shows the time dependences of He, O, and Fe for impulsive SEP event 44 during July 2000. The upper panel shows abundance enhancements of elements (labeled by *Z*) *versus A/Q*. The *middle panels* show the spectra of O, Ne, and Si (*left*) and the Ne/O and Si/O ratios (*right*).

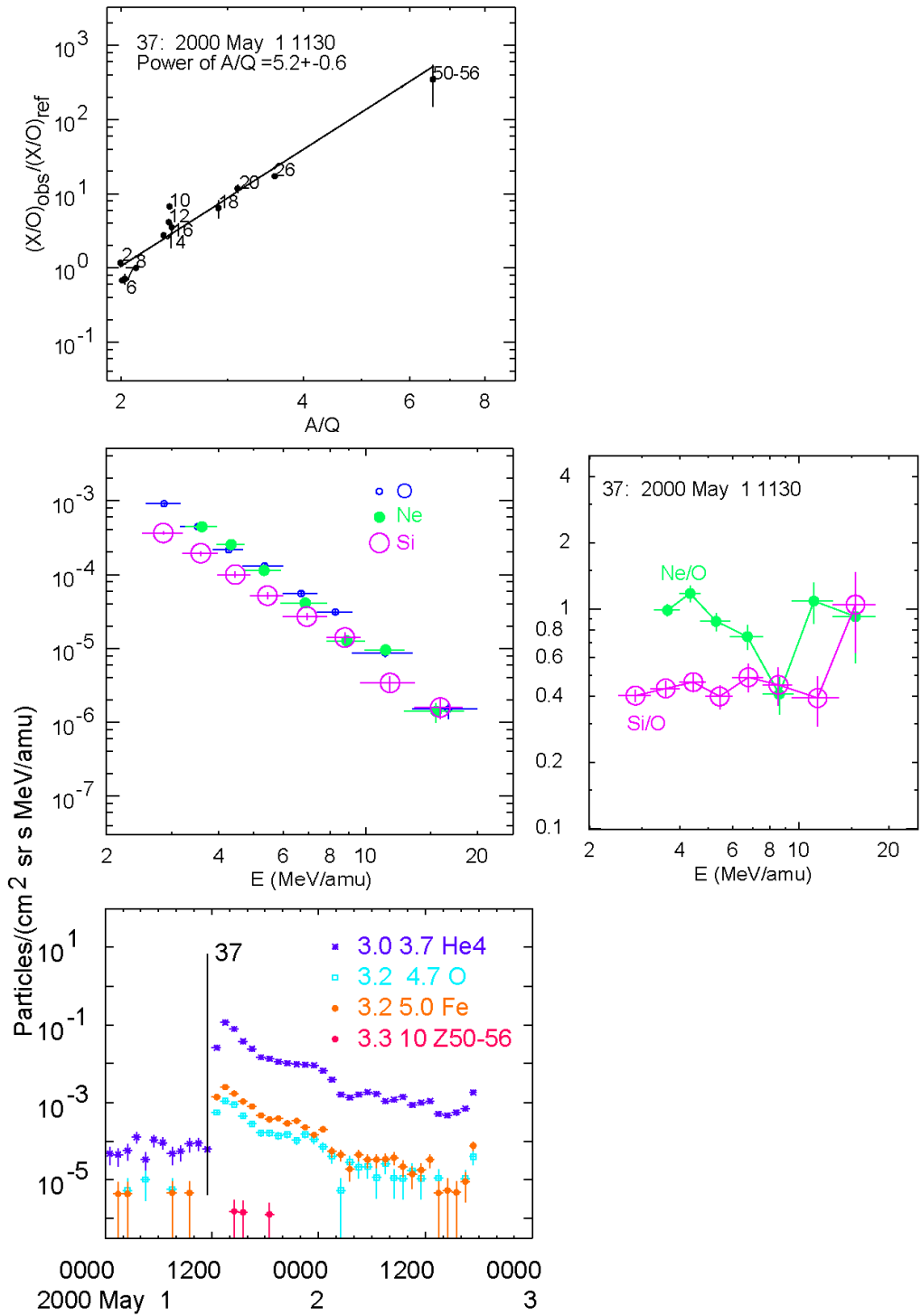

**Figure 12.** The *lower panel* shows the time dependences of He, O, Fe, and $50 \leq Z \leq 56$ for impulsive SEP event 37 beginning 1 May 2000. The upper panel shows abundance enhancements of elements (labeled by *Z*) versus *A/Q*. The *middle panels* show the spectra of O, Ne, and Si (*left*) and the Ne/O and Si/O ratios (*right*).

## 6. Discussion

The element abundance observations in impulsive SEP events lead to the following findings:

i) The relative suppression of He at the source can be an order of magnitude greater in impulsive SEP events (minimum source He/O ≈ 2) than in gradual SEP events (minimum source He/O ≈ 30). However, the He-poor events are uncommon and small and contribute little to residual impulsive suprathermal ions in the seed population for shock waves in gradual SEP events to sample.

ii) He/C often tends to decrease with energy in the He-poor events.

iii) Overall, except for the He-poor events, the mean and maximum values of source He/O are similar in impulsive and gradual SEP events.

iv) Suppression of He is unrelated to the *A/Q* dependence. Other element abundances (except possibly C) are completely unaffected by the large decrease in He. Essentially all of the events have *Q*-values consistent with source plasma temperatures of 2.5 – 3.2 MK.

v) Intensity variations along the energy spectra of all species are comparable with the 20 – 30% variations in abundances about the best-fit power law in *A/Q* for impulsive SEP events. Corresponding variations in gradual SEP events can be much smaller (Reames, 2016b, 2018b).

vi) An additional source of Ne with a steep spectrum contributing at low energies cannot be excluded.

It is not surprising that impulsive and gradual SEP events share much of the variation corresponding to $30 \leq$ He/O $\leq 100$. Both types of events sample similar regions of the solar corona and some gradual SEP events even reaccelerate residual ions from many impulsive SEP events. The suppression of He/O below a value of ≈ 90 is often discussed in terms of the uniquely high FIP of He = 24.6 eV which makes it especially slow to ionize in transit to the corona (Laming 2009, 2015; Reames 2018b). A strong suppression of He can be accompanied by smaller suppression of Ne and Ar (Laming 2009); it is possible to deduce small suppressions of Ne and Ar in the reference gradual SEPs relative to photospheric abundances (see Appendix A), but Ne is surely enhanced, not suppressed, in impulsive SEP events, even somewhat more than the power-law fits would suggest, while Ar is consistent with the power-law fits. Factor-of-two suppression of He has been considered theoretically (Laming 2009) while order-of-magnitude suppression is observed in He-poor events.

The rarer values of He/O < 30, especially He/O ≈ 2, that seem to be unique to impulsive SEP events, might be a property of the acceleration physics. This may also apply to the unique C-poor event 19 in Figure 10 which actually has measured C/O = 0.08 ± 0.04. C-poor events were previously observed below 1 MeV amu$^{-1}$ by Mason, Gloeckler, and Hovestadt (1979) and they were initially explained by assuming that the temperature and *A/Q* value of C prevented it from resonance enhancement in the wave-particle preheating mechanism that selected O and heavier ions (*e.g.* Fisk, 1978). However, while the Fisk (1978) model can suppress C, it cannot produce the strong power-law *A/Q* dependence of the elements from He to Pb which we now find (*e.g.* Figure 8 of Reames, Cliver, and Kahler, 2014a).

The more-recent acceleration model for $^3$He-rich events (Temerin and Roth, 1992; Roth and Temerin, 1997) has replaced the early preheating models which had become quite numerous. The Temerin-Roth model i) explains and uses the strong association observed between type III electrons and $^3$He-rich events (Reames, von Rosenvinge, and Lin, 1985; Reames and Stone, 1986), ii) it produces full acceleration of $^3$He and other ions, not just preheating, and iii) the streaming type-III electrons provide a self-consistent source for the resonant waves that preferentially energize $^3$He. However, this model does not explain the power-law dependence of the element abundances on *A/Q*, nor did its predecessors. Whatever the mechanism, the low value of $^4$He in the He-poor events does tend to magnify the $^3$He/$^4$He ratio by reducing the denominator.

The particle-in-cell simulation of islands of magnetic reconnection of Drake *et al.* (2009) does produce the power-law dependence on *A/Q* that is observed (but it does not explain the enhancement of $^3$He). This study also predicts the existence of a threshold value of *A/Q*, above which the enhancements occur. The threshold depends upon the plasma beta in the reconnection region (Drake *et al.*, 2009). Could the threshold fall just above *A/Q* = 2 so as to exclude He, or He and C, but not N and O, *etc.*, from enhancement? Such a sharp threshold seems unlikely and we have never seen N- or O-poor SEP events.  However, it is interesting that He and C are depressed by about the same factor in Figure 10.

Actually, individual ions cannot have values of $A/Q$ as near 2.00 as implied in Figure 1.  For example, a value of $A/Q = 2.006$ for C at 3.2 MK actually means that 98.5% of the $^{12}$C is $^{12}$C$^{+6}$ with $A/Q = 2.000$ and 1.5% is $^{12}$C$^{+5}$ with $A/Q = 2.400$.  Thus if He were depleted because it has $A/Q = 2$, 98.5% of the C would behave just like $^4$He$^{+2}$.  Also, at the same temperature, 57% of the O would be O$^{+8}$, so that C, N, and O should behave rather similarly.  Thus, it is extremely difficult for a threshold in $A/Q$ to suppress He without also affecting C, N, and O.  This is also true of other mechanisms that depress He (*e.g.* Steinacker *et al.*, 1997).

Direct measurements of ionization states of Fe in impulsive SEP events have shown evidence (DiFabio *et al.*, 2008) that electron stripping appears to increase with energy, suggesting that the ions have traversed small amounts of material after acceleration and attained equilibrium charge states.  However, the material traversal apparently is not enough to cause significant energy loss of the heaviest ions like Pb or to disrupt the power-law behavior *versus A/Q*.  In any case, material traversal of ions at $\approx 3$ MeV amu$^{-1}$ would certainly not decrease He/C; quite the contrary.  It has been known for many years that, for impulsive SEP events, the elements up through Si are fully ionized when they arrive near Earth (Luhn *et al.*, 1987).  They must be stripped after acceleration; otherwise it would be nearly impossible to produce the observed enhancements in Ne/O, Mg/O, and Si/O if all these ions have $A/Q = 2$.

For the He-poor events, we are left with the decrease in He/C with energy in nearly all of those events, and some decrease in He/C with energy is even seen in a few other impulsive SEP events (*e.g.* events 13 and 78 in Figure 5).  If He and C both have $A/Q \approx 2.00$, it is very difficult to explain why the spectrum of He is steeper than that of C.  Alternatively, He may be depressed in the corona because of its high FIP (Laming, 2009), but the strongly suppressed He-poor events and the C-poor events are still difficult to explain.

Mason *et al.* (2016) have observed impulsive SEP events below 1 MeV amu$^{-1}$ with a huge excess of S (*e.g.* S/O = 32).  Energy spectra of some elements in these events roll over at low energies, similar to the spectrum of $^3$He, and even peak at near 0.1 MeV

amu$^{-1}$. The authors suggest a low-temperature source plasma (0.4 MK) and possible resonant acceleration for some ions like that for $^3$He. The $^4$He intensities in these very small events are barely measurable in LEMT above 2 MeV amu$^{-1}$ and despite poor statistics they show no strong anomalies (S/O < 1). Thus these unusual events must have extremely steep spectra. Their relevance to the present work is that resonant acceleration of a single species, such as Ne, is possible and could be a contributor to the excess enhancement of low-energy Ne that we see. However, it is not at all obvious why Ne, with a nominal $A/Q \approx 2.5$, would be preferentially selected while nearby Mg with $A/Q \approx 2.45$ would not, but observed enhancement of a single species like S or Ne does suggest a resonant process.

The variations we see along the spectra or along the $A/Q$ power law may be easier to understand. Different islands of magnetic reconnection along a reconnecting region are likely to have different evolution, different source plasma and temperatures, producing different spectra and powers of $A/Q$. Ions that escape early will have different spectra from those that escape later. All of these effects combine to produce the abundances and spectra of a single event. Thus a single SEP event is an average over a microcosm that samples the same kind of event-to-event variations that are also seen in different SEP events. Impulsive SEP events are composites that probably involve source particles with a distribution of temperatures (around 2.5 – 3.2 MK) and that source distribution may change with the SEP energy we observe.

The relative abundance of hydrogen in impulsive SEP events is also interesting and is considered separately (Reames, 2019a). Hydrogen has also been considered in gradual SEP events as well (Reames 2019b).

**Acknowledgments:** The author thanks Chee Ng and Steve Kahler for helpful comments on this manuscript.

## Disclosure of Potential Conflicts of Interest

The author declares he has no conflicts of interest.

## Appendix A: Reference Abundances of Elements

The average element abundances in gradual SEP events are a measure of the coronal abundances sampled by SEP events (Reference gradual SEPs in Table 1). They differ from photospheric abundances (Table 1) by a factor which depends upon FIP (*e.g.* Reames 2018a; 2018b). Ion "enhancements" are defined as the observed abundance of a species, relative to O, divided by the reference abundance of that species, relative to O.

**Table 1** Reference element abundances are listed for each element, atomic number, and FIP, as abundances in the photosphere, in gradual SEPs, and in impulsive SEPs.

|       | Z     | FIP [eV] | Photosphere[1] | Reference Gradual SEPs[2] | Avg. Impulsive SEPs[3] |
|-------|-------|----------|----------------|---------------------------|------------------------|
| He    | 2     | 24.6     | 156000±7000    | 57000±5000                | 53000±3000[**]         |
| C     | 6     | 11.3     | 550±76[*]      | 420±10                    | 386±8                  |
| N     | 7     | 14.5     | 126±35[*]      | 128±8                     | 139±4                  |
| O     | 8     | 13.6     | 1000±161[*]    | 1000±10                   | 1000±10                |
| Ne    | 10    | 21.6     | 210±54         | 157±10                    | 478±24                 |
| Mg    | 12    | 7.6      | 64.5±9.5       | 178±4                     | 404±30                 |
| Si    | 14    | 8.2      | 61.6±9.1       | 151±4                     | 325±12                 |
| S     | 16    | 10.4     | 25.1±2.9[*]    | 25±2                      | 84±4                   |
| Ar    | 18    | 15.8     | 5.9±1.5        | 4.3±0.4                   | 34±2                   |
| Ca    | 20    | 6.1      | 4.0±0.7        | 11±1                      | 85±4                   |
| Fe    | 26    | 7.9      | 57.6±8.0[*]    | 131±6                     | 1170±48                |
| Zn-Zr | 30-40 | -        | -              | 0.04±0.01                 | 2.0±0.2                |
| Sn-Ba | 50-56 | -        | -              | 0.0066±0.001              | 2.0±2                  |
| Os-Pb | 76-82 | -        | -              | 0.0007±.0003              | 0.64±0.12              |

[1]Lodders, Palme, and Gail (2009), see also Asplund *et al.* (2009).

[*] Caffau *et al.* (2011).

[2] Reames (1995, 2014, 2017a).

[3] Reames, Cliver, and Kahler (2014a)

[**] This work.